\documentclass{INTERSPEECH2023}
\usepackage{hyperref}
\usepackage{adjustbox}
\usepackage{url}

\interspeechcameraready

\title{Differentially Private Adapters for Parameter Efficient Acoustic Modeling}

\name{Chun-Wei Ho$^1$,  Chao-Han Huck Yang$^1$, Sabato Marco Siniscalchi$^{1,2,3}$}
\address{
  $^1$Georgia Institute of Technology, U.S.A\\\
  $^2$Kore University of Enna, Italy \\
  $^3$Norwegian University of Science and Technology, Norway}
\email{\{chun-wei.ho,huckiyang\}@gatech.edu; marco.siniscalchi@ntnu.no}

\begin{document}

\maketitle
 
\begin{abstract}


In this work, we devise a parameter-efficient solution to bring differential privacy (DP) guarantees into adaptation of a cross-lingual speech classifier. We investigate a new frozen pre-trained adaptation framework for DP-preserving speech modeling without full model fine-tuning. First, we introduce a noisy teacher-student ensemble into a conventional adaptation scheme leveraging a frozen pre-trained acoustic model and attain superior performance than DP-based stochastic gradient descent (DPSGD). Next, we insert residual adapters (RA) between layers of the frozen pre-trained acoustic model. The RAs reduce training cost and time significantly with a negligible performance drop. Evaluated on the open-access Multilingual Spoken Words (MLSW) dataset, our solution reduces the number of trainable parameters by 97.5\% using the RAs with only a 4\%  performance drop with respect to fine-tuning the cross-lingual speech classifier while preserving  DP guarantees. 


\end{abstract}

\section{Introduction}

With the rapid growth of the computation ability and commercial datasets, more and more  personal data are collected, which poses the issue of protecting sensitive data. The United States Census Bureau, for instance, announced a new security standard \cite{bureau2021disclosure} based on Differential Privacy (DP) \cite{dwork2008differential}. The $(\epsilon, \delta)$-DP mechanism allows us to measure the security of algorithms and provides a guarantee based on a privacy budget. However, ensuring differential privacy degrades the system's performance \cite{bagdasaryan2019differential} because it restricts access to the data. In addition, training a large model with DP is not only time-consuming but also leads to a more severe drop in performance. \cite{bagdasaryan2019differential}.

Nonetheless, there are many benefits associated with the use of large-scale datasets and large models. For example, large-scale datasets are fundamental to deploying well-trained deep neural networks (DNNs) \cite{lecun2015deep, jean2017does}; moreover, if the size of the DNN is large enough, it can reach the global minima from any initialization with the gradient descent algorithm \cite{haeffele2015global}. Although the global optimality was only proven in tensor factorization, \cite{haeffele2015global} shows the benefits associated with large connectionist models. 
Indeed, there exist several large pre-trained models that have been proven vital for different downstream tasks \cite{devlin2018bert, baevski2020wav2vec, radford2021learning, raffel2020exploring, li2023efficient, hung2023low, chang2023speechprompt, yen2021neural} after fine-tuning - in this work, we will use the term fine-tuning and adaptation interchangeably. 


Unfortunately, fine-tuning a pre-trained larger model, in addition to being a time-intensive procedure, can also distort the pre-trained features and underperform out-of-distribution \cite{kumar2022fine}. Training large models with differential privacy is even harder because DP-related perturbations are introduced into the training process. Therefore, a feasible solution to estimate and exploit a representation of a large pre-trained model is becoming a pressing issue to be tackled. 


This work aims at investigating the benefits of leveraging model adaptation and parameter efficient techniques in the context of  differential privacy. In particular, we propose a cross-domain differential private fine-tuning framework \footnote{GitHub Link: \url{https://github.com/Chun-wei-Ho/Private-Speech-Adapter}.} leveraging a deep frozen model pre-trained on public source data, and private target data.  We consider the case when there is a domain mismatch between source and target domains. In the proposed framework the frozen pre-trained model doesn't guarantee privacy but provides information from non-sensitive source data. We also use additional parameters (weights) to serve as a domain adaptor, which provides information from the target data and introduces DP guarantees. In particular,  DP stochastic gradient descent (DPSGD) \cite{song2013stochastic, bassily2014private, abadi2016deep}, and Private Aggregation of Teacher Ensembles (PATE) \cite{papernot2016semi, yang2021pate} are used to attain DP guarantees.

\begin{figure}[tb]
    \centering
    \includegraphics[width=.5\textwidth]{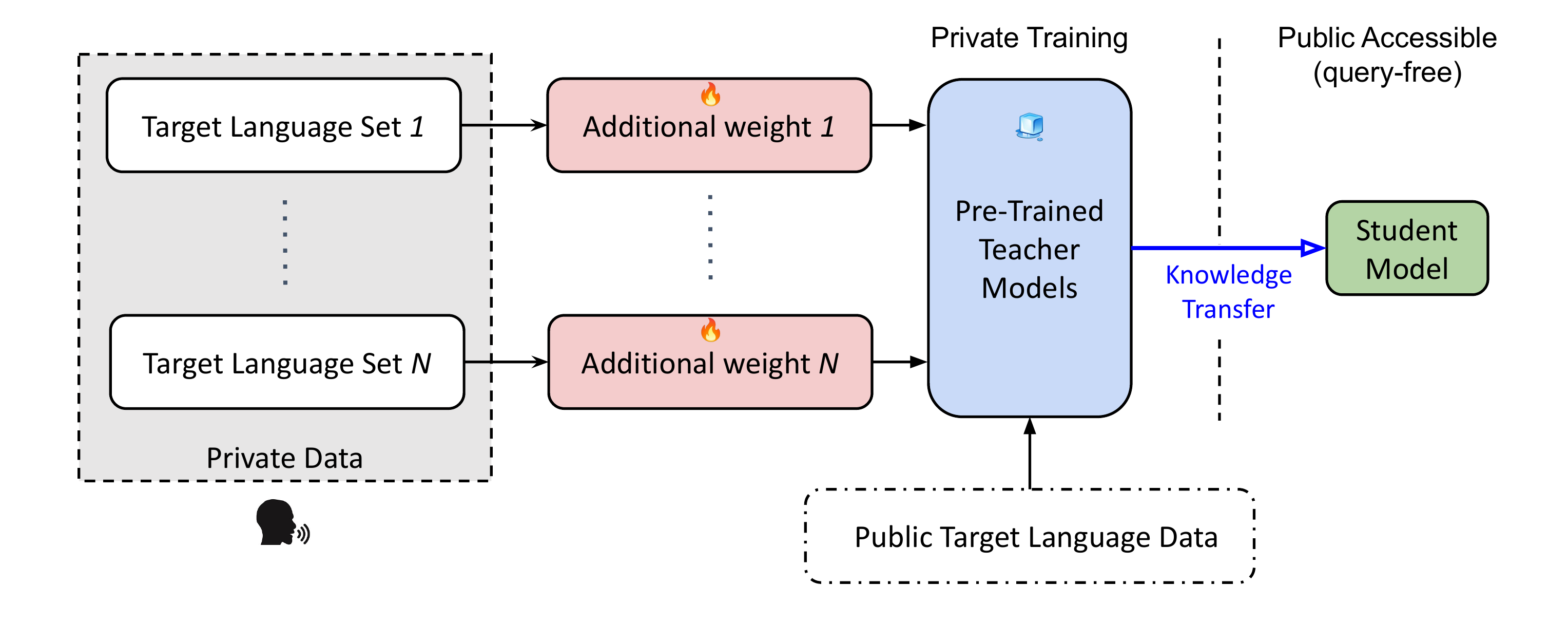}
    \caption{Proposed private aggregation of teacher ensembles~\cite{papernot2016semi} (PATE)-based adapter for parameter efficient fine-tuning on acoustic and speech processing.}
    \label{fig:pate_adapter}
\end{figure}

For DPSGD, we follow what was proposed by Da \etal in \cite{yu2021differentially}. Since the experimental evidence demonstrated poor results with DPSGD, we devised a PATE-based solution, which led to a substantial performance improvement. Figure \ref{fig:pate_adapter} shows the proposed PATE-based solution to perform model adaptation (fine-tuning) while attaining DP-privacy guarantees. The {\em additional weights} shown in the figure are trained on different disjoint chunks of the sensitive data. Those weights are then inserted into the frozen pre-trained large models using the solutions discussed in \cite{yu2021differentially}. The obtained frozen pre-trained model is aggregated with different weights together based on PATE's algorithm. Finally, the student model queries from the aggregated teacher model using non-sensitive target domain data and learns only from non-sensitive data to preserve privacy. To the best of the authors' knowledge, our work is the first to propose cross-domain DP-based  acoustic modeling adaptation.  The overall solution does not only guarantee DP, but it also is  parameter efficient.

%








\section{Related Works}

\label{section:doubleblind}

\subsection{Differential Privacy in a Nutshell}
\label{sec:dp}

The DP mechanism \cite{dwork2008differential} is established to evaluate the security of an algorithm. DP is parameterized by the  privacy budget variable   $\epsilon$, and $\delta$ defined as follows:

\textbf{Definition 1} An algorithm $\mathcal{A}$ is said to be $(\epsilon, \delta)$-DP if for all adjacent datasets $D$ and $D'$, and for any possible event $S$, the algorithm satisfies:
\begin{equation}
    \text{Pr}[\mathcal{A}(D) \in S] \leq e^\epsilon \text{Pr}[\mathcal{A}(D') \in S] + \delta
\end{equation}
The above equation, in some sense, guarantees that the outcomes of the algorithm with $D$ and $D'$ are indistinguishable.

There are several methods to achieve $(\epsilon$, $\delta)$-DP, and most of them require some DP-oriented perturbation. The perturbation guarantees $(\epsilon, \delta)$-DP by making the output of the algorithm, $\mathcal{A}(D)$ and $\mathcal{A}(D')$, indistinguishable. The simplest method to guarantee DP is to introduce the Laplace perturbation to the output of $\mathcal{A}$. It has been shown that we can achieve pure DP ($\delta=0$) with Laplace perturbation added \cite{sarathy2011evaluating}.

Although there exist several ways to estimate the privacy budget $\epsilon$, one of the most convenient methods is Renyi Differential Privacy (RDP) \cite{moore19_interspeech}, which is based on the Renyi Divergence by \eqref{eq:renyi-divergence}, which is similar to the Kullback-Leibler Divergence:
\begin{equation} \label{eq:renyi-divergence}
    D_\alpha(P \Vert Q)=\frac{1}{\alpha-1}E_{X\sim Q}\log \left (\frac{P(x)}{Q(x)} \right )^\alpha
\end{equation}
The RDP is  defined as follows:

\textbf{Definition 2} An algorithm $\mathcal{A}$ is said to be $\alpha, \epsilon$-RDP if for all adjacent datasets $D$ and $D'$, the algorithm satisfies:
\begin{equation}
    D_\alpha(f(D) \Vert f(D')) \leq \epsilon
\end{equation}

 It has been proven in \cite{moore19_interspeech} that if any algorithm satisfies $\alpha, \epsilon$-RDP, it's also an $\left(\epsilon+\frac{\log(1/\delta)}{\alpha-1}, \delta\right)$-DP algorithm. We use RDP to evaluate DP in this study.

\subsection{Privacy Preserving in Machine Learning}
\label{sec:DP-ML}
A common method to preserve privacy is by DP-based perturbations. However,  perturbations also degrade the system's performance. Finding a trade-off between  performance and privacy has become an important topic worth investigating. Two popular algorithms have been designed to preserve privacy in machine learning. The first is DP stochastic gradient descent (DPSGD) \cite{song2013stochastic, bassily2014private, abadi2016deep}, in which the effect of single data is restricted by per-utterance gradient clipping, and the noises are added to satisfy a certain privacy budget $\epsilon$. The second method is PATE \cite{papernot2016semi}, which is based on three stages: First, several teacher models are trained on disjoint chunks of  sensitive data. Then, the outputs of the teacher models $T_i(x, \theta_i)$ are aggregated using a private aggregation algorithm \eqref{eq:private_aggregation}. Finally, the student model is trained on some public data and the output of the teacher models, defined as  $T(x, \theta)$ in \eqref{eq:private_aggregation}. PATE models achieve $(\epsilon, \delta)$-DP by introducing noises in the aggregation phase and by hiding sensitive data from the student model. The amount of noise is determined by the ``smooth sensitivity'' \cite{nissim2007smooth} of the teacher models, which is also called data-dependent privacy. By reducing the required DP-oriented perturbation while aggregating, PATE has been tested as the state-of-the-art results in different applications, e.g., \cite{jordon2019pate,aslan2023price}.

\begin{equation} \label{eq:private_aggregation}
    T(x, \theta) = \text{argmax}\{T_i(x, \theta_i) + \text{Lap}_{i.i.d}(\lambda)\}
\end{equation}

\subsection{Parameter Efficiency \& Differential Privacy}
Training a huge deep model taking into account DP requirements can be troublesome because we have to restrict the information extracted from the data. Furthermore, the perturbation introduces randomness into the learning phase. The amount of perturbation required under the same privacy budget is depended on the model size. The larger the model is, the more perturbation we need to preserve privacy.  For example, the perturbation added to the gradients is proportional to the square root of the number of trainable parameters in DPSGD. That in turn leads to a trade-off between the model capacity, and DP guarantees. In many DP setups \cite{papernot2016semi, tramer2020differentially}, smaller and simpler model architectures end up providing superior performance. Nonetheless, Da \etal \cite{yu2021differentially} proposed to use parameter efficient methods to deal with the noise injection while training large models with DPSGD. In their study, it has been experimentally proven that larger models with parameter efficiency lead to better results when used in combination with DPSGD. We posit that parameter efficiency serves as a conduit between large models and privacy budgets. To this end, we investigate a first attempt to advance parameter-efficient learning with PATE, which has been demonstrated to have wide-ranging applications for performance-driven tasks.

\subsection{Parameter Efficient Algorithms}
\label{sec:method-lp-RA}
In this study, we mainly focus on two parameter efficient algorithms. Linear Probing (LP) \cite{kumar2022fine} prevents distortions by freezing the entire encoder while training the linear head\footnote{The last linear layer is referred to as ``head''} only. By reusing the pre-trained weights completely, Linear Probing is effective when the source domain and the target domain are similar to each other.

Adapters \cite{rebuffi2017learning} modifies the feature extractors by inserting some adapting layers without changing the pre-trained weights. More specifically, the relationship between the output of the $i^{th}$ layers $\hat{\mathcal{F}}_\theta^{i}(x)$ and the output of the $(i-1)^{th}$ layers $\hat{\mathcal{F}}_\theta^{i-1}(x)$ are described in \eqref{eq:residual_adapter}, where $\Theta$ denotes non-trainable parameters, and $\theta$ denotes trainable parameters. $\mathcal{A}_\theta$ denotes some non-linear function parameterized by $\theta$. The hat notation, $\hat{\cdot}$, indicates the functions whose inputs are the model input, $x$, instead of the output of the previous layer.

\begin{equation} \label{eq:residual_adapter}
    \begin{aligned}
    & \theta^*  = \arg \min_{\theta} \left \{  \mathcal{L}_{\text{error}}(\sigma(\hat{\mathcal{F}}_\theta^{N}(x)),\hat{y}) \right \}  \\
    & \text{where}~~~
    \begin{cases}
        \hat{\mathcal{A}}_\theta^i(x) =& \mathcal{A}_\theta^i(\hat{\mathcal{F}}_\theta^{i-1}(x))\\
        \hat{\mathcal{F}}_\theta^{i}(x) =& \underbrace{\mathcal{F}_\Theta^i (\hat{\mathcal{F}}_\theta^{i-1}(x))}_{\text{original encoder (frozen)}} + \underbrace{\hat{\mathcal{A}}_\theta^i(x)}_{\text{Adapter output}}
    \end{cases}
    \end{aligned}
\end{equation}

DNN Residule Adapter ($\text{RA}_\text{DNN}$) \cite{tomanek2021residual}, a common adapter uses a simple up-projector and a simple down-projector along with a residual path to define the non-linear function $\mathcal{A}_\theta$, which modifies the input feature, $\hat{\mathcal{F}}_\theta^{i-1}(x)$, by a limited matrix rank. It has been experimentally proven that $\text{RA}_\text{DNN}$ can attain comparable performance results to those obtained through a fine-tuning of the whole model parameters but using only  up to 2 \% of parameters \cite{houlsby2019parameter}.


\label{section:preprints}

\section{Proposed DP based Parameter Efficient Adaptation for Acoustic Modeling}

In this study, two of the most popular privacy-preserving algorithms, DPSGD and PATE were investigated. For DPSGD, we used the same setup in \cite{yu2021differentially}, where only the $\text{RA}_\text{DNN}$s are updated during training. Figure \ref{fig:pate_adapter} shows instead the proposed PATE-based solution, where $N$ different additional weights are trained on the different disjoint chunks from the sensitive dataset. The weights are then inserted into the global teacher model and are aggregated together using the private aggregation algorithm proposed in \cite{papernot2016semi}. The student model, on the other hand, learns from the public data queried from the private teacher models. Therefore,  the student can learn from private data without direct access to it. As explained in Section \ref{sec:DP-ML}, the amount of required DP-oriented perturbation is determined by the sensitivity of the teacher models. Therefore, by applying data-dependant privacy and domain adaptation, we were able to successfully reduce the amount of DP-oriented perturbation required to preserve privacy.


\subsection{DNN Residual Adapters Connection}
\label{sec:adapter_connection}

As discussed in Section \ref{sec:method-lp-RA}, $\text{RA}_\text{DNN}$ is one of the common parameter-efficient adapters. In this study, we also investigated different non-linear functions, $\mathcal{\hat{A}}_\theta(x)$. Inspired by \cite{yang2023english, yang2022rep}, we try to connect the $\text{RA}_\text{DNN}$s using some skip connections. Instead of just performing neighboring connections, we tried to connect the $\text{RA}_\text{DNN}$s in three different ways and investigate their effects. The three connection ways are summarized in \eqref{eq:connectoin}. The connections are inspired by Unet \cite{ronneberger2015u} and DenseNet \cite{huang2017densely}.

\begin{equation} \label{eq:connectoin}
    \begin{aligned}
        &\text{Neighboring:} \quad \hat{\mathcal{A}}_\theta^i(x) = \mathcal{A}_\theta^i(\hat{\mathcal{F}}_\theta^{i-1}(x)+\hat{\mathcal{A}}_\theta^{i-1}(x)) \\
        &\text{Unet-alike \cite{ronneberger2015u}:} \quad \hat{\mathcal{A}}_\theta^i(x) = \mathcal{A}_\theta^i(\hat{\mathcal{F}}_\theta^{i-1}(x)+\hat{\mathcal{A}}_\theta^{N-i}(x)) \ \forall i > \frac{N}{2} \\
        &\text{DenseNet-alike \cite{huang2017densely}:} \quad \hat{\mathcal{A}}_\theta^i(x) = \mathcal{A}_\theta^i(\hat{\mathcal{F}}_\theta^{i-1}(x)+\sum_{k=1}^{i-1}\hat{\mathcal{A}}_\theta^k(x))
    \end{aligned}
\end{equation}

As defined in \eqref{eq:connectoin}, the neighboring connections connect the output of the previous layer. In the Unet-alike connection, the last $i$ layers are connected to the first $i$ layers. And in the DenseNet-alike connection, every layer is connected to every preceding layer.

\subsection{Evaluation of Utility}
We leveraged Eric Hulburd's work \cite{hulburd2020exploring} to assess the quality of the proposed approach and used the utility defined in \eqref{eq:utility} that takes both  parameter efficiency and performance:

\begin{equation} \label{eq:utility}
    \text{Utility}=\frac{\text{Accuracy}-50}{\log(\text{Number of trainable parameters})}
\end{equation}




\section{Experiments \& Results}
\subsection{Experimental Setup}
We assessed our framework on a keyword classification task. Specifically, we used the English Google Speech Command V2 (EGSP-V2) \cite{warden2018speech} as  source domain, and the Multilingual Spoken Words \cite{mazumder2021multilingual} as  the target domain. We took into account only four languages, namely  English, German, French, and Russian, and generated smaller subsets from them, referred to as \textit{MLSW-mini} \footnote{The list of train/test split is reported on \url{https://github.com/Chun-wei-Ho/Private-Speech-Adapter}.}, to simulate low-resource conditions. \textit{MLSW-mini} configuration is shown in Table \ref{tab:MLSW-mini}. And the EGSP-V2 was used to pre-train the deep classifier. Then, we adapted the model to \textit{MLSW-mini} with DP. For DPSGD, we used \textit{MLSW-mini-train} and half of \textit{MLSW-mini-test} to train the model. For PATE, we trained the teacher models on \textit{MLSW-mini-train}. Then we trained the student model on half of the \textit{MLSW-mini-test}. The remaining data in \textit{MLSW-mini-test} was used for evaluation. The proposed setup follows the standard  PATE setup \cite{papernot2016semi}. The privacy budget $\epsilon$ is 8.0 \footnote{We follow a common privacy budget ($\epsilon$=8) based on \cite{yu2021differentially}~and Apple's official document in \url{https://www.apple.com/privacy/docs/Differential_Privacy_Overview.pdf}} for French, German, and English, and 11.6 for Russian.


\begin{table}[tb]\footnotesize
    \centering
    \caption{MLSW-mini dataset. "\# Words" indicates the number of unique words in the language. The sample rate of the waveforms is 16 kHz. Each waveform  is roughly 1 second long.}
    \vspace{-1mm}
    \label{tab:MLSW-mini}
    \scalebox{0.85}{
    \begin{tabular}{p{2 cm} c c c}
        \toprule
        Language & \# Words & \# Samples/word & Total Train Audio Time \\
        \midrule
        en (Germanic) & 18 & 4501-4927 & 23 hours 34 mins \\
        de (Germanic) & 15 & 4011-4910 & 18 hours 14 mins \\
        fr (Romance) & 13 & 4081-4988 & 16 hours 01 mins \\
        ru (Slavic) & 23 & 1002-4758 & 11 hours 00 mins \\
        \bottomrule
    \end{tabular}
    }
\vspace{-2mm}
\end{table}

The deep architecture used as a pre-trained model is the Keyword Transformer (KWT) \cite{berg2021keyword}. KWT first performs a time-distributed linear project of the mel-spectrogram; it  then concatenates the features with token embeddings. Next, the concatenated features are fed into 12 layers of transformer blocks with dimension 192 and classified using a linear head. The setups are similar to \cite{berg2021keyword} with the only difference being that  12 trainable $\text{RA}_\text{DNN}$s (with different dimensions) were inserted between the transformer blocks in the fine-tuning phase.

The Mel-spectrogram  generated with a 30 ms analysis window, a 10 ms frame shift, and 40-points DFT is the input feature used in both pre-training, and fine-tuning. For optimization, we used AdamW except for DPSGD. The number of epochs was set to 200. All the other setups are the same as those in \cite{berg2021keyword}.

\subsection{Cross-lingual Adaptation Results}

In this section, we investigate the effect of domain adaption with cross-lingual data and compare the two introduced DP algorithms, DPSGD and PATE, with and without $\text{RA}_\text{DNN}$s.

In Table \ref{tab:pate-dpsgd}, the baseline method, i.e., adapting the whole KWT network parameters without DP guarantees, attains a classification accuracy equal to 96.49 with a utility of 3.0 on the France language. By comparing the results with or without DP in Table \ref{tab:pate-dpsgd}, we can see that both utility and accuracy drop when DP constraints are imposed. In particular, the accuracy drops from 96.49 to 53.40 when DPSGD, the fourth row, is used, and the utility  drops from 3.0 to 0.22. PATE, in the fifth row, instead can limit the drop in accuracy and utility. 

Furthermore, by comparing the results of training from scratch (fs) and fine-tuning (FT), we conclude that domain adaptation is required to successfully train a model with DP. But differ from what was reported on language modeling in \cite{yu2021differentially}, DPSGD is not effective in cross-lingual acoustic adaptation but PATE is. We argue the difference is mainly because the domain mismatch is larger in cross-lingual tasks, and the mechanism of data dependant privacy in PATE reduces the amount of perturbation needed to be added under the same level of privacy budget. We also evaluate all the selected languages listed in Table \ref{tab:MLSW-mini}, reporting an overall average results in the last two rows in Table \ref{tab:pate-dpsgd}. The results indicate that our method works not only on French but also in multi-lingual scenario. 

\begin{table}[tb]\footnotesize
    \centering
    \caption{Comparison between DPSGD and PATE with and without $\text{RA}_\text{DNN}$s on \textit{MLSW-mini}. The All language results are the weighted average accuracy based on the number of utterances in the four selected languages.}
    \vspace{-1mm}
    \label{tab:pate-dpsgd}
    \scalebox{0.8}{
    \begin{tabular}{c p{2.4 cm} c r c c}
        \toprule
        lang & Method & DP & \# Train Para. & Utility & Acc. (\%)\\
        \midrule
        fr & from Scratch (fS) & \xmark & 5.4 M (100 \%) & 2.88 & 94.58 \\
        & fS w/ PATE  & \cmark & 5.4 M (100 \%) & 2.66 & 91.23 \\
        \hline
        en $\to$ fr & Fine-tune (FT) & \xmark & 5.4 M (100 \%) & 3.00 & 96.49 \\
        & FT w/ DPSGD & \cmark & 5.4 M (100 \%) & 0.22 & 53.40 \\
        & FT w/ PATE & \cmark & 5.4 M (100 \%) & 2.72 & 92.10 \\
        \cline{2-6}
        & LP w/ PATE & \cmark & 21.7 K (0.4 \%) & 1.13 & 61.13 \\
        & $\text{RA}_\text{DNN}$ w/ DPSGD  & \cmark & 0.9 M (14.6 \%) & 0.77 & 60.69 \\
        & $\text{RA}_\text{DNN}$ w/ PATE  & \cmark & 0.9 M  (14.6 \%) & \textbf{3.05} & 91.82 \\
        \hline
        all & fS & \xmark & 5.4 M (100 \%) & 2.71  & 92.03 \\
        & fS w/ PATE & \cmark & 5.4 M (100 \%) & 2.32 & 85.97 \\
        \hline
        en $\to$ all & FT & \xmark & 5.4 M (100 \%) & 2.99 & 96.38 \\
        & FT w/ PATE & \cmark & 5.4 M (100 \%) & 2.49 & 88.51 \\
        & $\text{RA}_\text{DNN}$ w/ PATE & \cmark & 0.9 M (17 \%) & \textbf{2.75} & 87.72 \\
        \bottomrule
    \end{tabular}
    }
    \vspace{-1mm}
\end{table}

\subsection{Residual Adapter Size Effect on Fine-tuning}
In this section, the effects of $\text{RA}_\text{DNN}$s are discussed. The \textit{MLSW-mini} French in Table \ref{tab:MLSW-mini} is used for our experiment. We performed the experiments with a privacy budget $\epsilon=7.96$ using a PATE~\cite{papernot2016semi} based KWT~\cite{berg2021keyword} model. We also used use $\text{RA}_\text{DNN-d}$ to denote that the down-projection dimension of the $\text{RA}_\text{DNN}$s is $d$. We tried several $d$ values ranging from 3 to 288, where 288 is twice the dimension of the original feature dimension, and 3 is instead 64 times smaller than the original feature dimension. As summarized in Table \ref{tab:adapter_fr}, $\text{RA}_\text{DNN-24}$ attains the best utility, with an 88.08 \%  accuracy training 2.46 \% of parameters only. In addition, by appropriately choosing the size of $\text{RA}_\text{DNN}$s, $\text{RA}_\text{DNN-288}$ provides a result that is  comparable with that of the fully fine-tuned model.

\begin{table}[tb]\footnotesize
    \centering
    \caption{Results with $\text{RA}_\text{DNN}$ for PATE with different $\text{RA}_\text{DNN}$ dimension and a privacy budget $\epsilon=7.96$ on \textit{MLSW-mini} French. $\text{RA}_\text{DNN-d}$ means the down-projection dimension is $d$.}
    \vspace{-1mm}
    \label{tab:adapter_fr}
    \scalebox{0.9}{
    \begin{tabular}{p{1.9cm} c r c c}
        \toprule
        Method  & DP & \# Train Para. & Utility & Acc. (\%) \\
        \midrule
        FT & \xmark & 5.4 M (100\%) & 3.00 & 96.49 \\
        FT w/ PATE & \cmark & 5.4 M (100\%) & 2.72 & 92.10 \\
        \hline
        LP & \cmark & 21.7 K (0.4\%) & 1.13 & 61.13 \\
        $\text{RA}_\text{DNN-24}$ & \cmark & 0.1 M (2.5\%) & \textbf{3.22} & 88.08 \\
        $\text{RA}_\text{DNN-288}$ & \cmark & 1.4 M (20.2\%) & 2.98 & \textbf{92.07} \\
        \bottomrule
    \end{tabular}
    }
    \vspace{-1mm}
\end{table}

\begin{figure}[tb]
    \centering
    \subfloat[Accuracy]{
        \includegraphics[width=.25\textwidth]{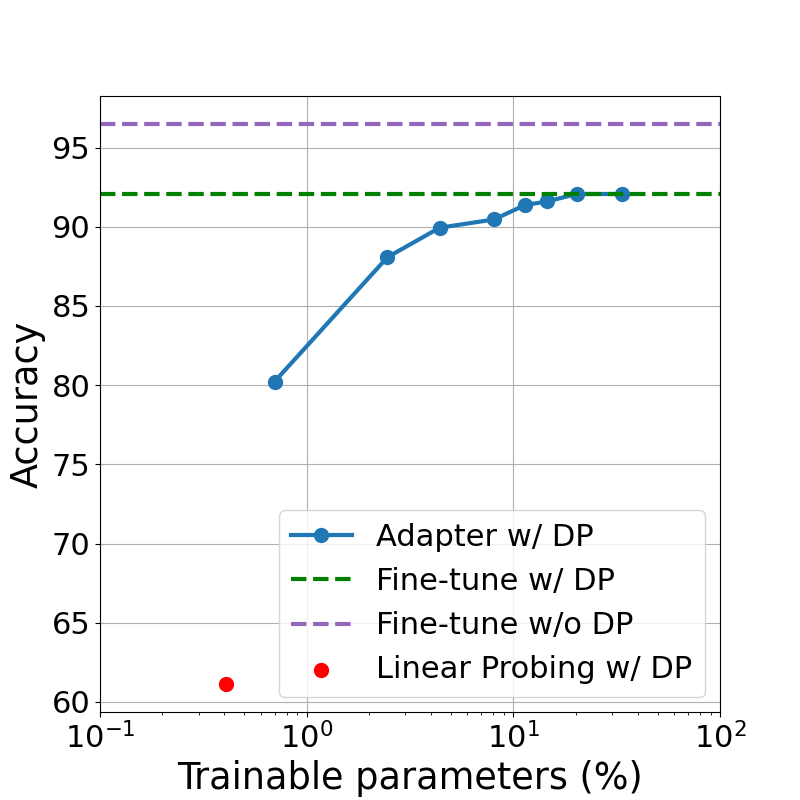}
    }
    \subfloat[Utility]{
        \includegraphics[width=.25\textwidth]{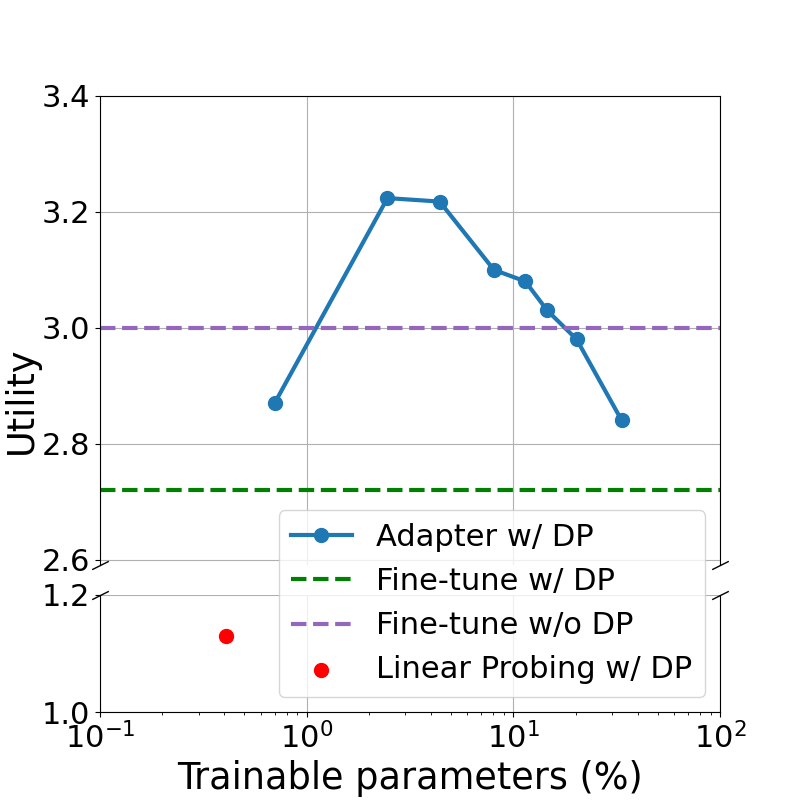}
    }
    \caption{Accuracy and utility of PATE-$\text{RA}_\text{DNN}$ architecture with different $\text{RA}_\text{DNN}$ sizes. (a) The model performance converges to the fine-tuning result when 20 \% of the parameters are adapted. (b) Our method achieves the best utility when 2.46 \% of parameters are adapted.}
    \label{fig:adapter-mlsw-fr}
\end{figure}

The effect of trainable parameters is also investigated in Figure \ref{fig:adapter-mlsw-fr}. First of all, as the number of trainable parameters increases, the model accuracy increases. However, it saturates at the fine-tuned accuracy when the number of parameters exceeds 20\% of the total model parameters. The latter means that we only have to train 20\% of the parameters to reach the best performance, and increasing the number of trainable parameters does not help. In addition, the best utility occurs when adapting  $2.46 \%$ of parameters. Reducing it does not lead to any beneficial effect, and the accuracy begins to degrade rapidly. Increasing the $\text{RA}_\text{DNN}$ size improves the overall accuracy, but the utility drops because the number of trainable parameters  increases accordingly.

\subsection{Different Connections of Residual Adapters}
We now investigate into the different $\text{RA}_\text{DNN}$ connections described in Section \ref{sec:adapter_connection}. As shown in Table \ref{tab:connection}, connecting the $\text{RA}_\text{DNN}$s in our task isn't necessarily helpful. We believe the reason is that the additional information from the other $\text{RA}_\text{DNN}$s is too noisy for a few-shot domain adaptation. We can validate the hypothesis from the fact that the DenseNet-alike connections provide the worst performance albeit it's more complicated than the other structures. And the results, same as our other experiments, lead to a conclusion that the simpler, the more promising.

\begin{table}[tb]\footnotesize
    \centering
    \caption{Experiments of PATE with different connections from EGSP-V2 to MLSW-mini French with $\epsilon=8.0$.}
    \vspace{-1mm}
    \label{tab:connection}
    \begin{tabular}{c p{2.5 cm} c}
        \toprule
        Model structure & Connection type & Acc. (\%) \\
        \midrule
        $\text{RA}_\text{DNN-24}$ & No connection & \textbf{88.08} \\
        & Neighboring~\cite{yang2023english} & 86.91 \\
        & Unet-alike & 87.61 \\
        & DenseNet-alike & 86.72 \\
        \hline
        $\text{RA}_\text{DNN-288}$ & No connection & \textbf{92.07} \\ 
        & Neighboring~\cite{yang2023english} & 91.49 \\
        & Unet-alike & 91.77 \\
        & DenseNet-alike & 91.13 \\
        \bottomrule
    \end{tabular}
    \vspace{-1mm}
\end{table}

\section{Conclusion}


In this work, we tackled the problem of preserving privacy in a  cross-lingual speech classification task. First, we tried to port what done on language modeling by \cite{yu2021differentially} using DPSGD, but we observed a significant performance drop using their method. Thus,  we proposed a novel PATE-based solution, which, differently from  DPSGD, led to a small drop in performance while still preserving DP guarantees.

Furthermore, to reduce the computational burden while fine-tuning with DP, we tested LP and $\text{RA}_\text{DNN}$. LP was not effective; whereas, $\text{RA}_\text{DNN}$ allows a reduction of 97.5\% of the parameters to be adapted while keeping a comparable performance of the PATE model. We also performed an ablation study to verify skip connection strategies on $\text{RA}_\text{DNN}$. Although skip-connection does not give any performance improvement, the exploring of different parameter-efficient architectures leveraging PATE is useful for future studies.\\
\textbf{Acknowledgments}
The authors would like to express their gratitude to Prof. Chin-Hui Lee from Georgia Tech for providing helpful insights and suggestions.


\clearpage
\bibliographystyle{IEEEtran}
\bibliography{mybib}




\end{document}